# A method for efficient measurement of gravitational lens time delays

**Gülay Gürkan**[*†]
*N. Jackson, I. W. A. Browne, L.V.E. Koopmans, C.D. Fassnacht, A. Berciano Alba*
*E-mail:* gulay.gurkan.g@gmail.com

The Hubble constant value is currently known to 10% accuracy unless assumptions are made for the cosmology [1]. Gravitational lens systems provide another probe of the Hubble constant using time delay measurements. However, current investigations of $\sim$ 20 time delay lenses, albeit of varying levels of sophistication, have resulted in different values of $H_o$ ranging from 50-80 km $s^{-1}$ $Mpc^{-1}$. In order to reduce uncertainties, more time delay measurements are essential together with better determined mass models [2, 3].
We propose a more efficient technique for measuring time delays which does not require regular monitoring with a high-resolution interferometer array. The method uses double image and long-axis quadruple lens systems in which the brighter component varies first and dominates the total flux density. Monitoring the total flux density with low-resolution but high sensitivity radio telescopes provides the variation of the brighter image and is used to trigger high-resolution observations which can then be used to see the variation in the fainter image.
We present simulations of this method together with a pilot project using the WSRT (Westerbork Radio Synthesis Telescope) to trigger VLA (Very Large Array) observations.
This new method is promising for measuring time delays because it uses relatively small amounts of time on high-resolution telescopes. This will be important because many SKA pathfinder telescopes, such as MeerKAT (Karoo Array Telescope) and ASKAP (Australian Square Kilometre Array Pathfinder), have high sensitivity but limited resolution.

[*] Speaker.
[†] I would like to thank TUBITAK for providing funds for my study.

## 1. An Efficient Method for Time Delay Measurements

It was demonstrated by [4] that gravitational lens time delays can be used to measure cosmological distances, hence the Hubble parameter $H_o$, assuming that the mass model of the lens potential is known. There are 19 gravitational lens systems which have time delay measurements among which only 5 lens systems have radio light curves and about 17 lenses have optical light curves. One reason for this is that there are fewer radio lenses that show significant variation. Additionally, lensing time-delay measurements require repeated imaging with a high-resolution (<1”) telescope to separate the images of a lensed source, and there are more optical telescopes than radio interferometer arrays that can conduct this. In the radio, VLA and MERLIN (Multi-Element Radio Linked Interferometer Network) are among the only interferometer arrays that are capable of regular imaging with the required resolution, which is expensive and disruptive [5]. For this reason we propose a technique which builds on a suggestion by [6], and minimizes the required time of high-resolution observations. We know from the theory of gravitational lensing that in asymmetric double and long-axis quadruple gravitational lenses the brighter image(s) shows variation first and after a short time the fainter component varies. Undertaking total flux monitoring gives us the variation of the brighter component, which dominates the total flux. It is therefore possible to use low-resolution but highly sensitive radio observations for total flux monitoring. As long as we detect the variation in the light curves, then around the time that we expect the fainter component to vary we can trigger high-resolution observations to see this variation in the fainter component’s light curve. Provided that $H_o$ is not too different ($\sim$15-20%) from 70 km s$^{-1}$ Mpc$^{-1}$, and that the lens galaxy’s mass profile is not too far from isothermal [7], then a reasonable period of followup should allow us to recover a time-delay using the high-resolution monitoring.

In order to assess the efficiency of our technique we performed simulations using the Pelt dispersion statistic and artificial light curves [8]. We evaluated the ability to recover a known time delay as a function of the number and separation of triggered observations. The results of the cross-correlation simulations demonstrated that time delays can be determined to $<$ 10% accuracy. For the simulations we used the light curve of the brighter component of a lensed source B1608+656 [9], obtained through high-resolution VLA observations and the light curve of the fainter component was simulated. From figure 1 it can be seen that we detect a minimum around 38 days which is within 10% of the actual time delay, 36 days.

## 2. Pilot Project

Total flux monitoring of 8 radio lens systems was conducted using WSRT observations with enough sensitivity to see the variation in the total flux. 5 GHz snap-shot observations with 3.7 arcsec resolution were collected in 39 epochs from 16 July to 30 October 2007. For the data reduction the NRAO (National Radio Astronomy Observatory) AIPS (Astronomical Image Processing Software) package was used. 3C147, a steep-spectrum source whose flux on WSRT scales is unlikely to vary significantly, was used as an absolute flux calibrator for all epochs. Initial editing was applied to each source before gain calibration. After all calibration, integrated flux densities of the target sources and the calibrators were derived by fitting a point-source model to the *(u,v)* data within the Caltech DIFMAP (Difference Mapping) software package.

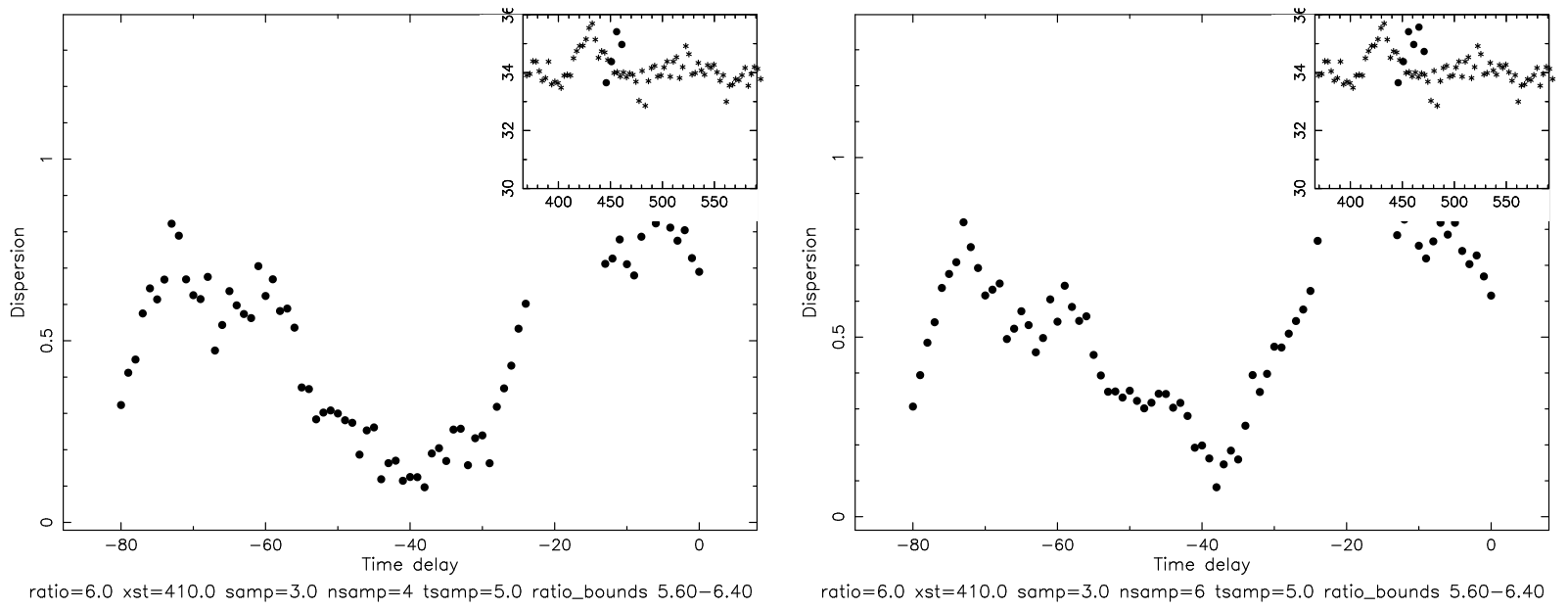

**Figure 1:** The Pelt statistic simulations were implemented using the data of double lens system B1608+656 [9]. The inset is a light curve of two components in simulated observations. The fluxes of the weaker component are scaled by the flux ratio and shifted by the actual time delay. A minimum in the dispersion statistic corresponds to a likely time delay. The two panels above show the results of triggered observations spaced 5 days apart (This was found to be the optimal spacing): (left) 4 samples of triggered observations cannot show a sharp minimum in the dispersion spectrum; (right) 6 triggered observations recovered a clean minimum around 38 days.

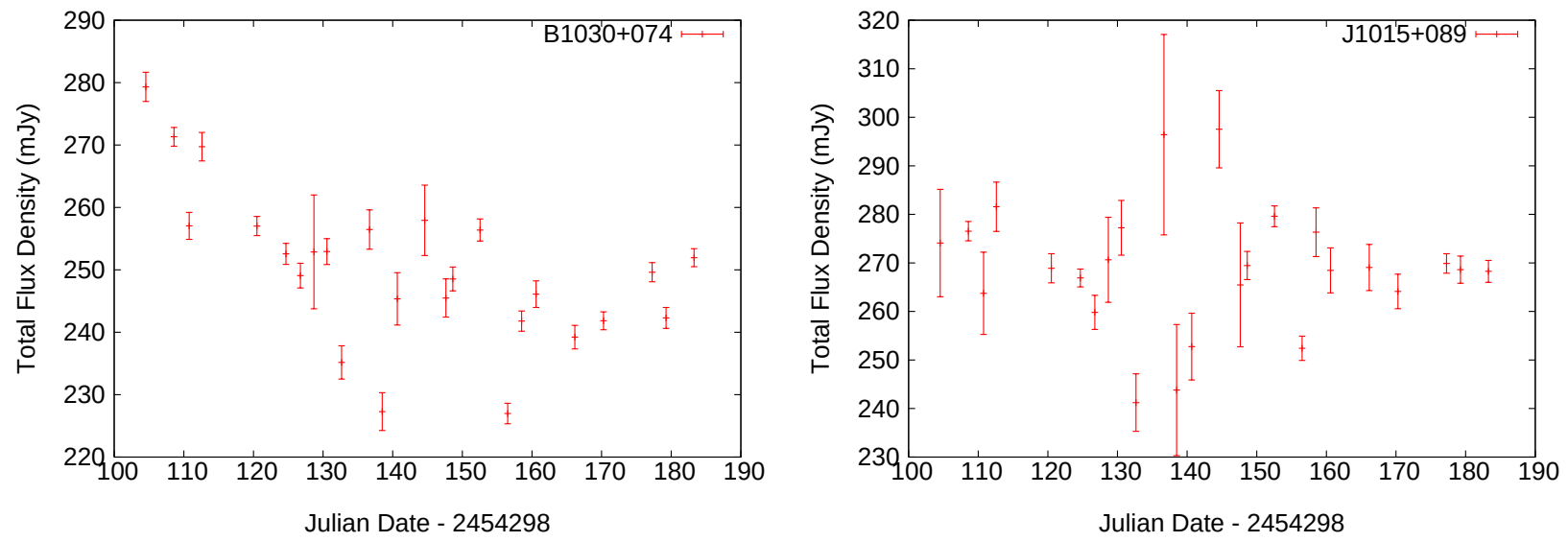

**Figure 2:** The WSRT light curves of B1030+074 and J1015+089.

Among the 8 systems only B1030+074 showed a variability during the total flux monitoring. It can be seen from figure 2 that there is a general trend of decreasing flux with time early in the observing period for the target source B1030+074. On the other hand, the flux density of the calibrator J1015+089 did not show the same trend.

VLA observations were carried out for B1030+074 between October 5, 2007 and January 3, 2008 in 9 epochs. The VLA observing period was chosen in order to catch the likely variation in the fainter component for any value of $H_o$, 50<$H_o$<100 km s$^{-1}$ Mpc$^{-1}$, assuming the lens galaxy has a isothermal mass profile. The data were collected at 5 GHz using 2 IFs each with 50 MHz bandwidth, in A configuration giving 0.4 arc-sec resolution. For the amplitude calibration, 3C286 was used as a primary calibrator and 1143+185, a compact source that does not vary [10], was used as a secondary calibrator to improve errors in the absolute flux calibration for each epoch. 1015+089 and 1014+088 were used as phase calibrators in different epochs.

## 3. Results

We could not catch the variation of the fainter image during the EVLA monitoring. How-ever, further cross-correlation simulations may provide time limits for the time delay between the images. (Gürkan et al. 2011 in prep.)

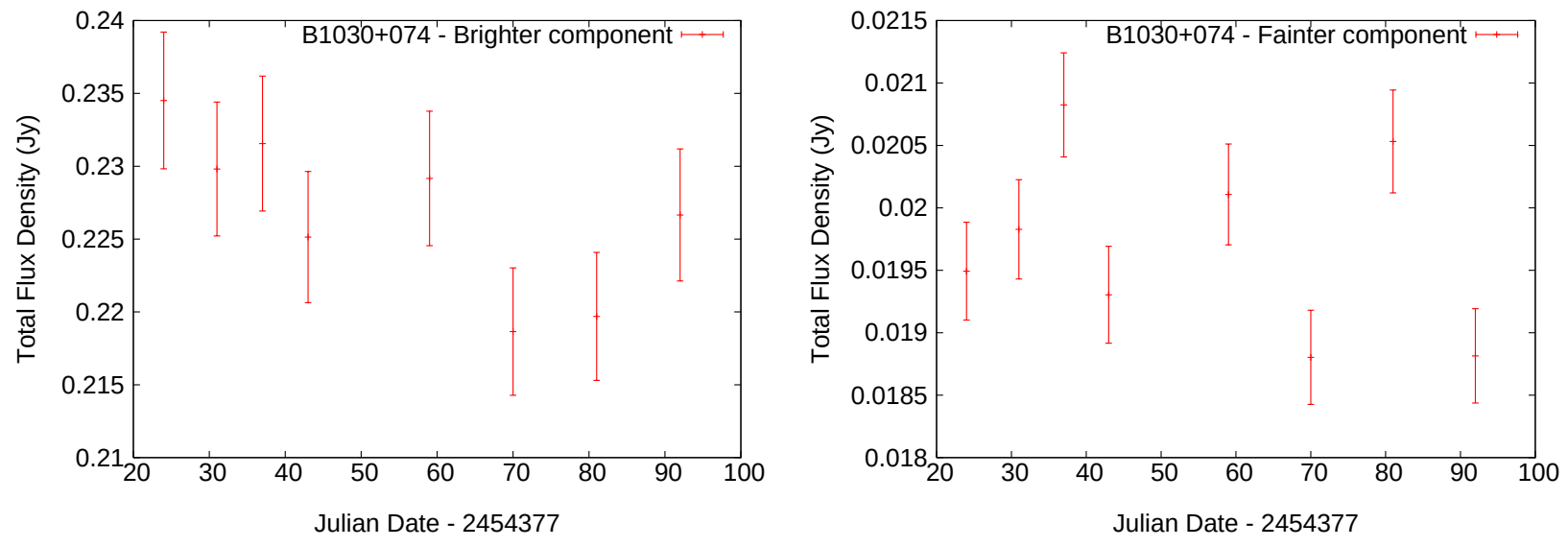

**Figure 3:** VLA light curves of the brighter and fainter components of B1030+074.

This new method is potentially useful because it predominantly uses time on low-resolution telescopes which have more time easily available. This is important because new, highly sensitive but low-resolution, instruments are under construction such as MeerKAT and ASKAP. Such instruments, together with a modest amount of high-resolution observational followup, may in future be useful for gravitational lensing time delay measurements by means of our new method.